\documentclass[doublecol]{epl2} 

\title{Photoabsorption spectra and the X-ray edge problem in graphene}

\shorttitle{Photoabsorption spectra in graphene}

\author{Georg R\"oder\inst{1} \and Grigory Tkachov\inst{2} \and Martina Hentschel\inst{1}}
\shortauthor{R\"oder, Tkachov, Hentschel}

\institute
{                    
  \inst{1} Max-Planck-Institut f\"ur Physik komplexer Systeme, N\"othnitzer Str. 38, 01187 Dresden, Germany\\
  \inst{2} Fakult\"at f\"ur Physik und Astronomie and R\"ontgen Center for Complex Material Systems, Universit\"at W\"urzburg, Am Hubland, 97074 W\"urzburg, Germany
}


\abstract{
We study the photoabsorption cross section and Fermi-edge singularities (FES) in graphene. 
For fillings below one half, we find, besides the expected FES in form of a peaked edge at the threshold (Fermi) energy, 
a second singularity to arise at excitation energies that correspond to the Dirac point in the density of states. 
We can explain this behaviour by comparing our results with the photoabsorption cross section of a metal 
with a small central band gap where we find a very similar signature. 
The existence of the second singularity might prove useful for an experimental determination of the Dirac point. 
We also demonstrate that the photoabsorption signal is enhanced by the zigzag edge states due to their metallic-like character. 
Since the presence of the edge states indicates a topological defect at the boundary,  
our study gives an example for a Fermi-edge singularity in a system 
with a topologically nontrivial electronic spectrum.
}

\begin{document}

\maketitle

\date{\today}



%
%

\section{Introduction}

The new carbon material - graphene \cite{geim} - exhibits metallic-type conduction and is  optically transparent at the same time. 
This unique combination of its electronic and optical properties has its origin in the two-dimensional nature of the material and its band structure 
that contains conical (Dirac-like) degeneracy points. Various condensed matter manifestations of graphene's unusual spectrum have been studied in recent years \cite{Neto09}. 
However, most of these studies deal with low excitation energies that probe the quasiparticle states on the Dirac cones in the vicinity of the Fermi level. 
For this reason and in view of potential applications, it is essential to investigate the response of graphene also to high energy perturbations, 
e.g. such that eject carriers from inner electronic shells. This occurs, for instance, in the X-ray excitation of a 
core electron into the conduction band. In conventional metals such an excitation is associated with a singularity at the threshold (Fermi) energy 
in the photoabsorption spectra, which is referred to as Fermi-edge singularities (FES) in the X-ray edge problem \cite{ohtaka:rmp}. 
In the present paper we study the analogue of the X-ray edge problem in graphene. 

One contribution to the photoabsorption cross section near the threshold frequency $\omega_{\mathrm{th}}$ comes from 
Anderson's orthogonality catastrophe (AOC) \cite{Anderson:prl}. In graphene AOC was found to be suppressed at half filling (i.e., at the Dirac point (DP)) \cite{aoc_graph}, 
a behaviour that can influence the Kondo effect and FES in graphene as was studied, e.g., in Ref.~\cite{xray_graph}. 
Here we consider FES in a more general situation, where AOC competes with a second, counteracting many-body response 
known as Mahan's exciton or the Mahan-Nozi{\`e}res-DeDominicis response \cite{Mahan, Nozieres3} (see, also Ref.~\cite{ohtaka:rmp}). 
In particular we consider the equivalent of the peaked-edge situation in metals (see below).
We show that the vanishing density of states (DOS) at the DP yields photoabsorption spectra that are very similar to those in a gapped material.
The existence of an additional clear photoabsorption signature associated with the DP could therefore be used for its experimental identification. 

We also examine the influence of graphene edges and the accompanying edge states on the FES. 
The photoabsorption signal is found to be significantly enhanced in zigzag-terminated graphene due to the presence of gapless edge states \cite{Fujita96}.  
They form a Kramer's pair of counter-propagating states along the same boundary and decay exponentially into the interior. 
The spectral and transport manifestations of such edge states and their connection 
to the Dirac physics have been discussed extensively in recent years (e.g. \cite{Waka00,Peres06,Brey06,GT07,GT09,TB,Akhmerov08,Basko09,GT_MH09,GT_EH10}).
In particular, there is a striking similarity between the graphene edge states and those found  
in two-dimensional topological insulators~\cite{Kane05,Bernevig06,Koenig07,QSH}. 
In this sense, the zigzag edge can also be viewed as a topologically nontrivial extended defect, due to which  
the FES in this system differs qualitatively from that in conventional bounded mesoscopic systems, e.g. ballistic quantum dots \cite{fes_prl, fes_prb, georg_fes}. 
In the latter systems the deviations from the bulk FES  have been attributed to the finite number of particles, the presence of mesoscopic fluctuations 
leading to both a broad distribution of Anderson overlaps and the photoabsorption cross section, and, most importantly, 
to modifications of the dipole matrix element compared to the metallic case that result 
from the self-interference of the wave function in the confined geometry\cite{fes_prl, georg_fes}.

\begin{figure}[tb]
\begin{center}
\includegraphics[width=1.\columnwidth]{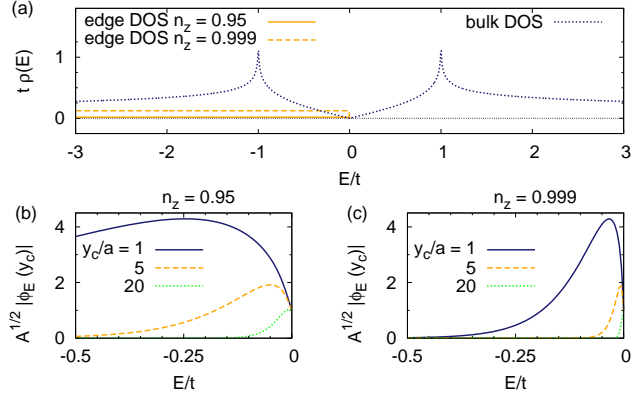}	
\end{center}
\caption{(Color online) 
(a) Bulk (dotted) and edge DOS (dashed and solid) $\rho(E)$ in graphene. 
The edge DOS is calculated for the zigzag-type boundary condition characterized by the parameter $n_z$ (see, also text). 
Note that edge states contribute only below the Dirac point as was confirmed in tunneling experiments~\cite{Koba05,Niimi06}. 
(b), (c) Local edge-state amplitude $\phi_E (y_c)$ normalized by sample area $A = L W$ versus energy at different distances $y_c$ from the boundary, 
measured in units of the interatomic distance $a$ = 0.142 nm ($W/a=50$). 
For $n_z \to 1$, the edge state amplitude has its maximum near the DP (see, panel c). 
}
\label{fig1}
\end{figure}

The paper is outlined as follows. We next give a brief introduction to the model and method that we apply to compute the photoabsorption cross section. The discussion of the results starts with bulk graphene. Then we proceed by considering the edge states on a generalized zigzag boundary and, in the end, summarize our findings.


 
\section{Model and method}

Our starting point is the nearest-neighbor tight-binding model for  
a rectangular graphene flake of length $L$ (along the $x$-axis) and width $W$ (along the $y$-axis, total area $A=L W $). 
The spectral properties of graphene are characterized by the DOS per one spin projection, containing both the bulk and edge contributions, 
shown in Fig.~\ref{fig1}(a), for the energy $\epsilon = E/t$ ranging from -3 to 3 in units of the nearest-neighbor 
tight-binding hopping parameter $t$=2.7 eV. 
The global edge DOS is calculated for one zigzag-type boundary by exact integration of the local DOS over the distance from the edge  
in a semi-infinite geometry \cite{GT_MH09}, with the $x$ and $y$ axes being parallel and perpendicular to the zigzag edge, respectively.
The model involves a single parameter $n_z$ that characterizes the degree of particle-hole asymmetry of the edge spectrum, resulting from the instability of the flat zero-energy edge band~\cite{Fujita96}.  
Near the two inequivalent valleys $\pm$ of graphene's Brillouin zone the edge-state spectrum has the following form~\cite{Akhmerov08,GT_MH09}:
\begin{eqnarray}
 & E_+(k)=-\hbar v_* k\,\,\, (k\geq 0),\quad E_-(k)=\hbar v_* k \,\,\, (k\leq 0),&
\label{spectrum}\\
&v_*=v\sqrt{1-n^2_z} < v,&
\label{v_*}
\end{eqnarray}
where $v_*$ is the edge-state velocity ($v$ is the bulk Fermi velocity at half-filling). We note that the deviation from the ideal zigzag boundary (i.e. from $n_z=1$) shifts the edge states below $\epsilon=0$, which is consistent 
with the tunneling spectra observed in experiments \cite{Koba05,Niimi06}.  
The dependence of the edge-state velocity $v_*$ (\ref{v_*}) on $n_z$ implies that 
the level spacing and, hence, the number of edge states in a given energy window  
is controlled by this parameter. 

The local edge-state wave function amplitude $\phi_E(y_c,n_z)$ depends on three parameters: energy $E$ ($\epsilon = E/t$, respectively), distance from the edge $y_c$ and parameter $n_z$ [see also Figs.~\ref{fig1} (b) and (c)]: 
\begin{equation}
\phi_E (y_c, n_z) =
    \frac{\sqrt{\frac{4}{3} \frac{n_z}{n_x} \,|\epsilon|}  \exp{\left(-\frac{2}{3} \frac{n_z}{n_x} \, \frac{y_c}{a} \, |\epsilon|\right)} }
    {\sqrt{La} \sqrt{1-\exp{\left(-\frac{4}{3} \frac{n_z}{n_x} \, \frac{W}{a} |\epsilon|\right)}}},
    \label{edgestate_yc}
\end{equation}
where $n_x = \sqrt{1-n_z^2}$. Evidently, the closer the edge to the ideal zigzag boundary (i.e. $n_z \to 1$) the more localized is the edge state at the boundary. For $n_z$=0.999 there is essentially no penetration of edge DOS into the graphene bulk.  
Note that we assume the bulk states to have uniform amplitudes (normalized to one)  throughout the sample
\cite{aoc_graph}, in contrast to the position ($y_c$)-dependent edge state wave function amplitude, Eq.~(\ref{edgestate_yc}).

For later use we define the filling parameter $f$ as the portion of the energy band that is filled, i.e., 
\begin{equation}
f = \frac{\epsilon_{\mathrm {filled}}^{\mathrm{max}} + 3}{6}.
\end{equation}
Edge states, when present, are included in this definition.

{\it Photoabsorption cross section.}
Our calculation of the photoabsorption cross section is based on the Golden-rule approach to the X-ray edge problem \cite{ohtaka:rmp,fes_prl,fes_prb}. 
We model the perturbation associated with
the excitation of a core electron into the conduction band, causing the FES, as a localized, rank-one perturbation \cite{ohtaka:rmp,kostya}. Its strength is scaled by the mean level spacing, and the perturbation is thus measured in terms of the dimensionless parameter $v_n$, see Refs.~\cite{aoc_graph, fes_prb} for details. 
The photoabsorption cross section 
depends then only on the perturbed and unperturbed energy levels and wave function amplitudes at the position $y_c$ of the perturbation. Note that in the general mesoscopic situation \cite{fes_prl,fes_prb} 
the fluctuations of the energy levels and, in addition, the non-uniform, position-dependent amplitudes result in considerable fluctuations of the photoabsorption cross section.

The photoabsorption cross section depends crucially on the dipole matrix element, in particular on the fulfillment of the dipole selection rules. 
Given the $p$-character of the electrons in the graphene band containing the Fermi level that is subject of our consideration here, 
dipole selection rules are fulfilled with $s$-type core electrons. 
This situation is referred to as $K$-edge in metals \cite{citrin,ohtaka:rmp}, 
and the corresponding dipole matrix element is proportional to the wave function amplitude. 
Note the distinct difference to the metallic case where, assuming the usual $s$-type conduction electrons, 
the dipole selection rules are fulfilled at the $L$-edge ($p$-type core electron), 
implying a contribution from the so-called Mahan-Nozi{\`e}res-DeDominicis response that overcompensates the AOC response, 
causing this edge to be typically peaked \cite{Mahan,ohtaka:rmp}.  
In contrast, the $K$-edge, where AOC is the only many-body effect contributing to the FES in the X-ray problem, is typically rounded \cite{ohtaka:rmp, citrin}. 
Here we focus exclusively on a situation that is equivalent to the metallic $L$-edge (dipole selection rules fulfilled) -- 
in graphene it is realized with an $s$-type core electron.

\begin{figure*}[t]
\begin{center}
	\includegraphics[width=0.8\linewidth]{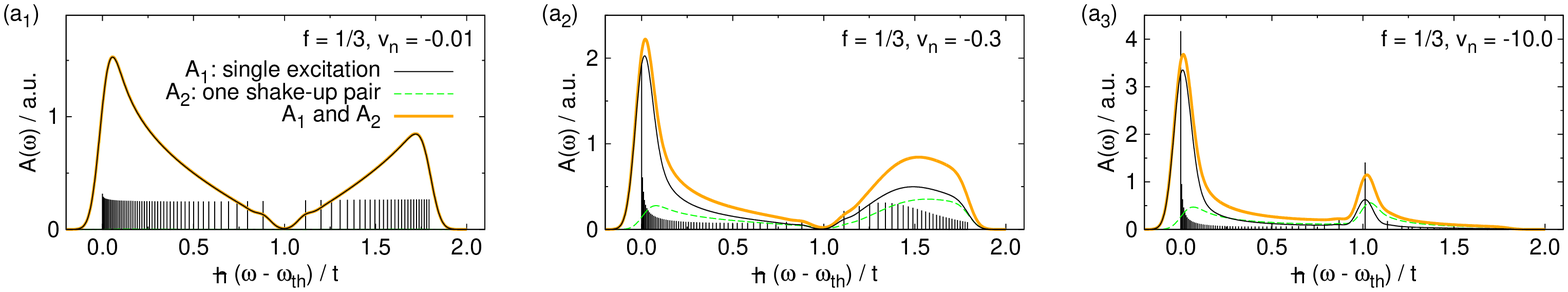}
  \includegraphics[width=0.8\linewidth]{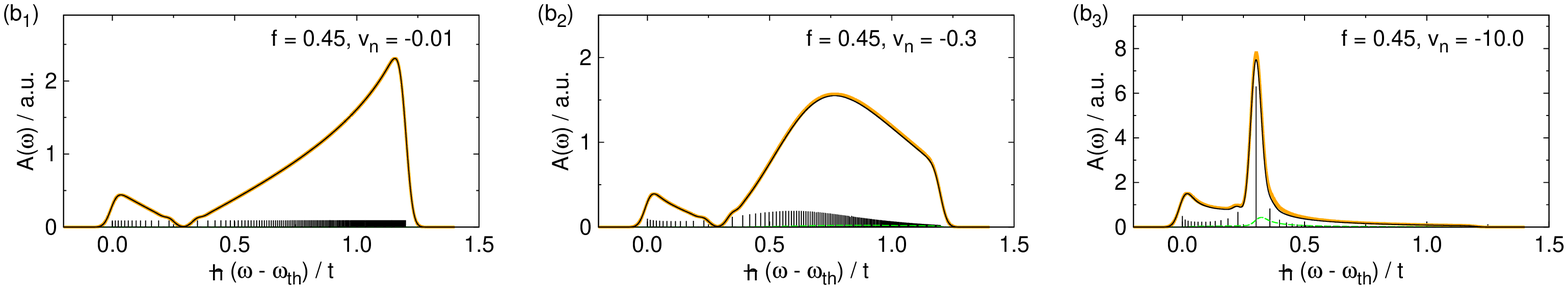}
  \includegraphics[width=0.8\linewidth]{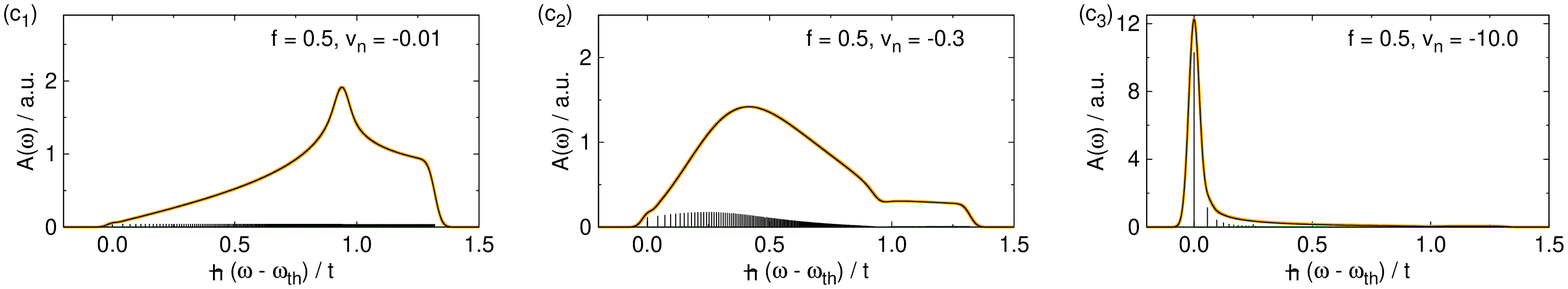}
  \includegraphics[width=0.8\linewidth]{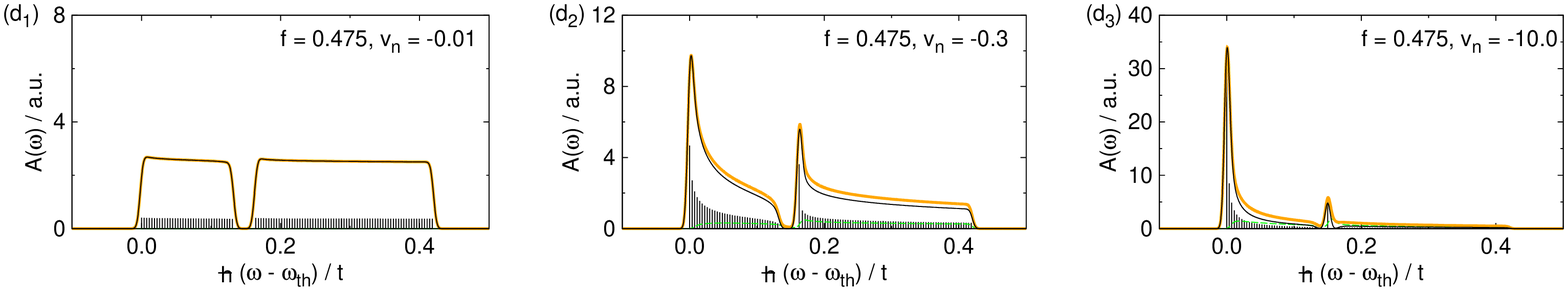}
\end{center}  
	\caption{(Color online) 
Photoabsorption spectra in clean/bulk graphene with an armchair boundary without edge states. 
The same behaviour can be found, independent from the boundary conditions chosen, in the center of (big) graphene flakes. 
The panels in the upper three rows, (a1) to (c3), refer to bulk graphene and show the total photoabsorption cross section (yellow/grey thick solid line, 
normalized such that the area underneath is one) consisting of the contribution of so-called direct and replacement processes 
($A_1$, black solid line) and processes involving the formation of one shake-up pair ($A_2$, green dashed line). 
The vertical lines illustrate that the photoabsorption cross section originates from discrete-energy events (arbitrary units). 
The perturbation strength increases from left to right, the filling from top to bottom. Note that, for small perturbations, 
the photoabsorption resembles the graphene (bulk) DOS; the fall-off of the photoabsorption signal on the right is unphysical and 
indicates the maximum excitation energies $\hbar \omega$ (measured w.r.t. the threshold energy $\hbar \omega_{\mathrm th}$ in units of $t$) considered. 
For large perturbations and fillings below 1/2, the existence of an additional singularity beyond the Fermi edge and associated with the DP, is evident, 
see in particular (a3) and (b3). The lower panels (d1) to (d3) were obtained for the toy model with a constant DOS and a central band gap 
(band from -3 to 3 as before, gap from -0.015 to 0.015). Total number of states $N=400$ in (a1)  to (a3), otherwise $N=1600$. }
\label{fig2}
\end{figure*}

\begin{figure*}[t]
\begin{center}
\includegraphics[width=0.8\linewidth,clip]{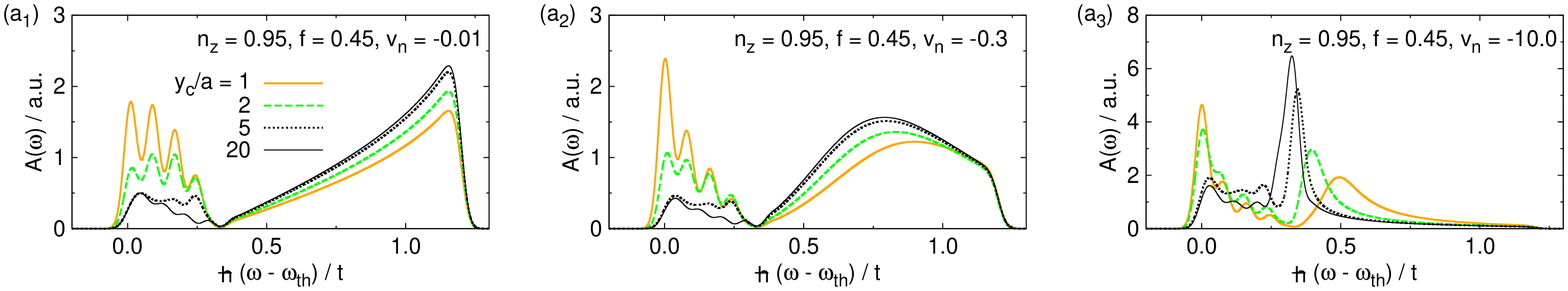}
\includegraphics[width=0.8\linewidth]{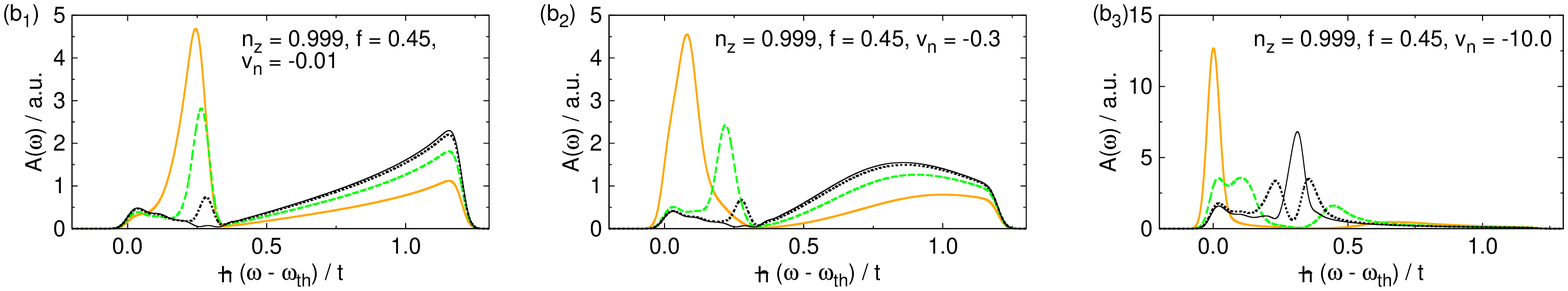}
\includegraphics[width=0.8\linewidth]{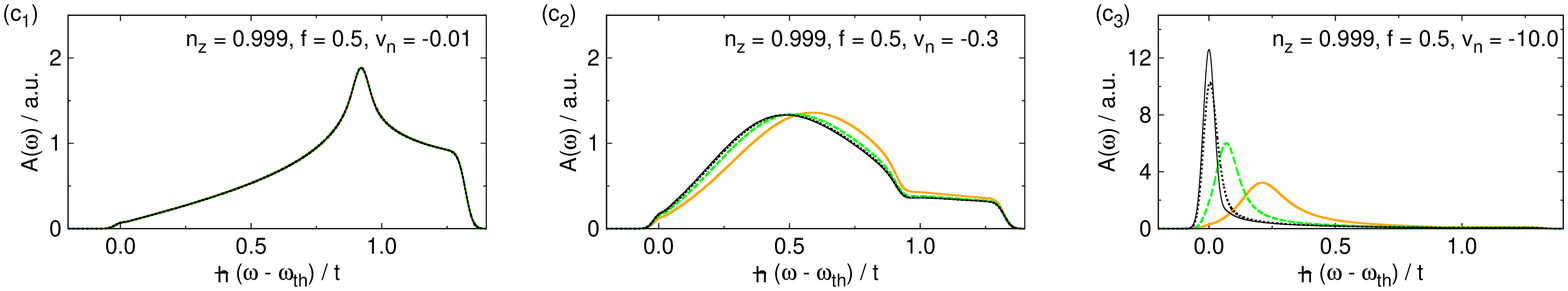}
\end{center}
\caption{(Color online) Total photoabsorption cross section in the presence of edge states taken at varying distance $y_c$ of the perturbation from the zigzag boundary. All curves are normalized such that the area underneath is one. The filling $f$ and the edge state parameter $n_z$ are varied; the perturbation strength increases from left to right. Note that the photoabsorption in panels (b1) to (b3), for the highest $y_c/a = 20$, closely follows the bulk photoabsorption, cf.~Fig.~\ref{fig2}(b1) to (b3), i.e., edge states play no role towards the center of the sample. The role of edge states for the particular shape of the FES becomes clear when looking at the curves $y_c = 1,2,5$; in particular, the weight of the first (conventional) FES increases towards the system boundary, 
whereas the weight of the second singularity, at the DP, decreases at the same time (and is shifted to larger enegies). 
}
\label{fig3}
\end{figure*}

The sudden appearance of a core hole triggers a non-adiabatic many-body response of the system that is in particular characterized by the excitation of electron-hole pairs from part of the excitation energy introduced into the system. However, consideration of one shake-up pair processes was shown to capture practically all of their contribution to the photoabsorption cross section \cite{fes_prb}. In the following, we will therefore discuss only the zero- and one-shake-up pair contribution to the photoabsorption cross section, marked as $A_1$ and $A_2$, respectively, and denoted by the black solid and (green) dashed line, respectively, in Figs.~\ref{fig2} and \ref{fig3}. We also neglect intervalley scattering because the interaction with the core hole has purely electrostatic character that does not involve the pseudospin degrees of freedom. Concerning the calculation of the photoabsorption cross section, we closely follow the method described in detail in Ref.~\cite{ohtaka:rmp} and recently applied to mesoscopic quantum dots \cite{fes_prl,fes_prb,georg_fes,japan}. We refer the interested reader to these papers as well as to Ref.~\cite{georg_diss}.




\section{Results 1: Bulk graphene}

We start the discussion of our results with the case where no additional edge states are present, a situation that is realized, e.g., in clean graphene with armchair boundaries, or in the center of (big) graphene flakes where the edge-state intensity does not contribute any more. 

The results for the photoabsorption cross section are shown in Fig.~\ref{fig2} for different fillings - $1/3$ (0.45 and 0.5) in the first (second, third) row of panels, respectively. The perturbation is increased from left to right. The very weak perturbation of $v_n=-0.01$ (left column) reveals, as expected, just the (almost unperturbed) DOS via the photoabsorption cross section. Note that the photoabsorption cross section is cut on the right due to the finite excitation energies taken into account. Shown are the contributions $A_1$ without shake-up pairs (so-called direct and replacement processes \cite{ohtaka:rmp}, black vertical and solid lines) and with one shake-up pair ($A_2$, green dashed curve). The yellow/grey (thicker) solid line is the total photoabsorption cross section, obtained as sum $A_1 + A_2$.

For intermediate and large $|v_n|$ (central and right column), the photoabsortion cross section deviates from the DOS and a FES in form of a threshold peak develops 
at the Fermi energy threshold. This behaviour is similar to that found in metals in the corresponding situation where the dipole selection rule is fulfilled ($L$-edge) 
and indicates the importance of the Mahan-Nozi{\`e}res-DeDominicis contribution. 
The half-filled case, panels (c1) to (c3) deserves an extra discussion: Note the absence of the FES at the threshold energy for the intermediate perturbation strength, 
in agreement with Refs.~\cite{xray_graph, aoc_graph}, due to the vanishing DOS at the DP. 
This behaviour contrasts the metallic (Fermi-liquid) signature that is, however, recovered for fillings below (and similarly, above) the DP, cf.~panels (a2) and (b2), 
again in agreement with Refs.~\cite{xray_graph, aoc_graph}. 
The existence of a clear peak in the photoabsorption even at half-filling in the case of a very strong perturbations, panel (c3), 
is a consequence of the formation of a bound state \cite{ohtaka:rmp} and of the (replacement) processes involving it \cite{fes_prb}. 
It also illustrates that true many-body effects (involving the interaction of the core hole with all conduction electrons) are less important in such situations. 

Interestingly, for fillings below one half, an additional singularity appears in the case of strong perturbations when the excitation energy reaches the DP. Its relative height w.r.t.~the Fermi threshold FES depends on the filling $f$ and it can actually be higher than the FES peak at the Fermi threshold. We point out that the shake-up contribution to the photoabsorption cross section carries more weight at the second singularity 
which is easily understood by the larger excitation energy and the related increase in the number of processes involving (one) shake-up pairs.

We now further investigate this, 
on first sight surprising, behaviour of developing an additional singularity for fillings below one half above the threshold energy that is, to a certain extent, reminiscent of the so-called opening of a second band in metals \cite{ohtaka:rmp}. In metals, it involves the formation of a bound state, i.e., it requires strong perturbations. Note that this applies also to the graphene case as the second singularity is developed best for the strongest perturbations (right panels).

We compare our results to that of a toy model consisting of a metal with a constant DOS and a band gap, cf.~Fig.~\ref{fig2}(d1) to (d3). The DOS ranges again from -3 to 3, with a small central band gap ranging from -0.015 to 0.015, cf.~the DOS-like photoabsorption for $v_n=-0.01$. As the perturbation strength $|v_n|$ is increased, clearly a second FES peak develops when the excitation energies reaches the onset of the second band, very similar to the situation encountered in graphene at the DP and in particular for fillings well below one half, see, e.g., panels (a3) and (d3).

We point out that the existence of the second singularity at the DP for fillings below one half provides the possibility to actually experimentally detect the DP via a photoabsorption measurement (results for more practicable transport measurements will be reported elsewhere).

%


\section{Results 2: Graphene with edge states}
We now include the presence of the edge states in the DOS and investigate their influence on the photoabsorption cross section, cf.~Fig.~\ref{fig3}. To this end we consider, as for the bulk graphene, different perturbation strengths and fillings, as well as two different values of the edge parameter $n_z$, namely $n_z=0.999$, describing an almost ideal zigzag edge, and, more generally, $n_z=0.95$ that allows for a particle-hole asymmetry of the edge states.

Figures~\ref{fig3}(a1) -- (a3) show the photoabsorption cross section for $n_z = 0.95$ and a filling of 0.45. 
For relatively weak perturbation [see Fig.~\ref{fig3}(a1)] the edge-state contribution is reminiscent of the energy dependence of the discrete local DOS: 
It has a multiple-peak structure below the DP, which is suppressed with increasing 
distance from the boundary (cf. curves for $y_c/a=1$ and $y_c/a=20$).  
For intermediate and large perturbation strengths [see Figs.~\ref{fig3}(a2) and (a3)], the edge-state contribution becomes more singular: 
Close to the boundary ($y_c/a=1$), a peaked FES with superimposed edge state ``wiggles'' develops at and near the Fermi edge. 
Further away from the boundary, i.e., closer to the center of the system ($y_c/a=20$), $A(\omega)$ is determined by the bulk DOS 
whose contribution is stronger at the second singularity that develops, as before, at the DP. 
Thus, the singularity associated with the DP is most pronounced if the core electron is excited away from the graphene boundary. 
Upon approaching the edge the second peak becomes smaller and shifts to somewhat larger energies [see light (yellow) curve for $y_c/a=1$ in Fig.~\ref{fig3}(a3)].   
Note that, with more discrete edge states in a given energy interval, their DOS becomes more homogeneous, resulting in denser wiggles in the photoabsorption cross section. 

Figures~\ref{fig3}(b1) -- (b3) show the photoabsorption cross section for an almost ideal zigzag edge with $n_z = 0.999$ and the filling of 0.45.  
Since for $n_z\to 1$ the edge-state dispersion [see Eq.~(\ref{spectrum})] gradually transforms into a flat band with vanishing velocity $v_*\to 0$, 
the edge states are localized much closer to the boundary, and their intensity and the local DOS are much narrower in energy [see Fig.~\ref{fig1}(c) 
and Eq. (\ref{edgestate_yc})]. Consequently, for relatively weak perturbation [Fig.~\ref{fig3}(b1)] the photoabsorption cross section displays the edge-state peak 
just below the DP, which depends very strongly on distance from the edge $y_c$.
As the perturbation strength $v_n$ is increased to $|v_n|=10$ [Fig.~\ref{fig3}(b3)], the peak shifts towards the Fermi threshold energy. 
The second peak near the DP is, in contrast, not well developed and can only be observed if the position of the perturbation is located away from the boundary ($y_c/a = 20$). 

For the filling of $1/2$,  there are no edge states above the Fermi energy as it now coincides with the DP.  
In this case the photoabsorption spectra are qualitatively similar to the bulk graphene [cf. panels c1-c3 in Figs.~\ref{fig2} and  \ref{fig3}].


\section{Conclusion}
We have studied the photoabsorption cross section and the X-ray edge problem in graphene flakes with different lattice terminations, allowing for  the presence or absence of edge states. Our results were obtained for spinless electrons at zero temperature and their generalization to higher temperatures is straightforward \cite{ohtaka_highertemp}.
Note that the photoemission spectra are then readily obtained by applying the Crooks relation \cite{crooks}.

In bulk graphene, i.e., far away from the boundaries of the graphene flake (or, in general, in the absence of edge states), we find a particularly interesting behaviour of the FES in the photoabsorption cross section. For fillings below $1/2$ (i.e., below the DP) and especially for strong perturbations, a second peaked singularity develops at the DP besides the well-known FES peak at the Fermi energy threshold. The studies of the photoabsorption spectra could, therefore, serve as an alternative means for the experimental identification of the DP. A comparison with the photoabsorption cross section of a metal with gapped DOS revealed a very similar behaviour and provided an understanding of this, at first sight, surprising behaviour in the sense that the suppressed DOS at the DP acts similar to a gapped DOS. 

In a graphene flake with a zigzag-like boundary the photoabsorption cross section is influenced by the edge states, especially for fillings close to $1/2$. As the edge-state intensity exponentially decreases from the boundary, they can actually ``see'' the localized potential associated with the core hole (left behind after the X-ray excitation of a core electron) only when this perturbation is located sufficiently close to the system boundary. Because of their metallic-like  
DOS the edge states enhance the FES, whereas the bulk graphene states provide the major contribution to the second, additional singularity that develops at the DP. 

\acknowledgments

M.H. thanks F. Guinea and E. Mucciolo for helpful discussions and the DFG for support in the Emmy Noether Programme and through the research group FG 760.


\begin{thebibliography}{99}
\bibitem{QSH}
However, the possibility of the quantum spin Hall state \cite{Kane05,Bernevig06,Koenig07} distinquishes 
the two-dimensional topological insulators in quality from graphene.
\end{thebibliography}
\end{document}